\pdfoutput=1

\documentclass[11pt]{article}

\usepackage[]{EMNLP2023}

\usepackage{times}
\usepackage{latexsym}

\usepackage[T1]{fontenc}

\usepackage[utf8]{inputenc}

\usepackage{microtype}

\usepackage{inconsolata}

\usepackage{amsmath,amsfonts}

\usepackage{supertabular,booktabs}
\usepackage{float}
\usepackage{supertabular}
\usepackage{subcaption}
\usepackage{makecell}
\usepackage{array}
\usepackage[symbol]{footmisc}
\usepackage{tabularx}
\usepackage{xcolor}
\usepackage[graphicx]{realboxes}
\usepackage{ragged2e} 
\newcolumntype{L}{>{\RaggedRight\hangindent=2em\hangafter=5}X}

\newsavebox{\mybox}

\usepackage{amsthm}
\usepackage{mdframed}
\usepackage{centernot}
\usepackage{comment}
\usepackage{hyperref}
\usepackage{graphicx}
\usepackage{booktabs}
\usepackage{float}

\newcommand{\Dd}{\mathcal{D}}

\newcommand\vd[2]{d_{i, p}}
\newcommand{\set}[1]{\left\{ #1 \right\}}

\newcommand{\R}{\mathbb R}

\newcommand{\Real}{\R}

\usepackage{tikz}

\definecolor{gold}{rgb}{0.99,0.78,0.07}
\usetikzlibrary{arrows,shapes,snakes,automata,backgrounds,positioning,decorations.pathmorphing,
	decorations.markings,calc}

\tikzstyle{dtreenode}=[draw=blue!10!gray,rounded rectangle, minimum size=5mm,fill=blue!10!white]
\tikzstyle{dtreeleaf}=[draw=black!60,minimum width=1cm,minimum height=0.4cm,rectangle,fill=blue!50!white]
\tikzset{every loop/.style={looseness=7}}
\tikzset{
	gluon/.style={decorate,draw=black,
		decoration={coil,amplitude=1pt, segment length=5pt}}
}
\tikzset{
	gluon1/.style={decorate,draw=black,
		decoration={coil,amplitude=3pt, segment length=3pt}}
}
\tikzset{
	gluonew/.style={decorate,draw=black,
		decoration={coil,amplitude=1pt, segment length=2pt}}
}

\tikzset{bicolor/.style args={#1 and #2 and #3}{
		path picture={
			\tikzset{rounded corners=0}
			\fill [#1] (path picture bounding box.south west)
			rectangle
			($(path picture  bounding box.north west)!#3!(path picture bounding
			box.north east)$);
			\fill [#2]
			($(path picture bounding box.south west)!#3!(path picture bounding
			box.south east)$)
			rectangle (path picture bounding box.north east);
}}}

\tikzset{tricolor/.style args={#1 and #2 and #3 and #4 and #5}{
		path picture={
			\tikzset{rounded corners=0}
			\fill [#1] (path picture bounding box.south west)
			rectangle
			($(path picture  bounding box.north west)!#4!(path picture bounding
			box.north east)$);
			\fill [#2]
			($(path picture bounding box.south west)!#4!(path picture bounding
			box.south east)$)
			rectangle
			($(path picture  bounding box.north west)!#5!(path picture bounding
			box.north east)$);
			\fill [#3]
			($(path picture bounding box.south west)!#5!(path picture bounding
			box.south east)$)
			rectangle (path picture bounding box.north east);
}}}

\usepackage[many]{tcolorbox}
\tcbuselibrary{listings,skins}

\lstdefinestyle{mystyle}{
  xleftmargin=0pt,
   basicstyle={\footnotesize\ttfamily},
   aboveskip=3mm,
   belowskip=3mm,
   keywordstyle=\bfseries,
   showstringspaces=false,
  escapechar=?,
  language=Java
}
\definecolor{code_indent}{HTML}{CCCCCC}

\newtcblisting{mylisting}[2][]{
  arc=0pt, outer arc=0pt,
  listing only,
  listing style=mystyle,
  title={\large #2},
  #1
}

 \definecolor{dkgreen}{rgb}{0,0.6,0}
 \definecolor{gray}{rgb}{0.5,0.5,0.5}
 \definecolor{mauve}{rgb}{0.58,0,0.82}

\definecolor{cadmiumgreen}{rgb}{0.0, 0.42, 0.24}
\definecolor{verde}{rgb}{0.25,0.5,0.35}
\definecolor{jpurple}{rgb}{0.5,0,0.35}
\definecolor{darkgreen}{rgb}{0.0, 0.2, 0.13}

\usepackage[ruled,vlined,linesnumbered,lined]{algorithm2e}
\usepackage{algpseudocode}

\usepackage{mathtools,xparse}

\usepackage{tabu}
\usepackage{multirow}

\usepackage{pifont}
\usepackage{enumerate}
\usepackage[shortlabels]{enumitem}

\usepackage{pgfplotstable}
\usepackage{pgfplots}

\usepackage[noframe]{showframe}
\usepackage{framed}

 {\endMakeFramed}
 \definecolor{shadecolor}{gray}{0.85}

\definecolor{bgblue}{RGB}{245,243,253}
\definecolor{ttblue}{RGB}{91,194,224}

\mdfdefinestyle{mystyle}{%
  rightline=true,
  innerleftmargin=10,
  innerrightmargin=10,
  outerlinewidth=3pt,
  topline=false,
  rightline=false,
  bottomline=false,
  skipabove=\topsep,
  skipbelow=false
}

\newtcolorbox{myboxi}[1][]{
  breakable,
  title=#1,
  colback=white,
  colbacktitle=white,
  coltitle=black,
  fonttitle=\bfseries,
  bottomrule=0pt,
  toprule=0pt,
  leftrule=3pt,
  rightrule=3pt,
  titlerule=0pt,
  arc=0pt,
  outer arc=0pt,
  colframe=black!50,
}

\newtcolorbox{myboxii}[1][style=mystyle]{
  breakable,
  freelance,
  colback=white,
  colbacktitle=white,
  coltitle=black,
  fonttitle=\bfseries,
  bottomrule=0pt,
  boxrule=0pt,
  colframe=white,
  after skip=0pt,
  overlay unbroken and first={
    \draw[white!75!black,line width=3pt]
    ([yshift=-9pt]frame.north west) --
    ([yshift=9pt]frame.south west);
  },
}
\title{On the Potential and Limitations of Few-Shot In-Context Learning to Generate Metamorphic Specifications for Tax Preparation Software}

\author{Dananjay Srinivas$^{*1}$ \;\;\; Rohan Das$^{*1}$ \;\;\; \textbf{Saeid Tizpaz-Niari}$^2$ \\ \textbf{Ashutosh Trivedi}$^1$ \;\;\; \textbf{Maria Leonor Pacheco}$^1$ \\
$^1$ University of Colorado Boulder \;\;\;
$^2$University of Texas El Paso \\
\texttt{$^1$\{first.last\}@colorado.edu} \;\;\; $^2$\texttt{saeid@utep.edu}}

\begin{document}
\maketitle

\footnotetext[1]{Equal Contribution}

\begin{abstract}
Due to the ever-increasing complexity of income tax laws in the United States, the number of US taxpayers filing their taxes using tax preparation software (henceforth, tax software) continues to increase. 
According to the U.S. Internal Revenue Service (IRS), in FY22, nearly $50\%$ of taxpayers filed their individual income taxes using tax software. 
Given the legal consequences of incorrectly filing taxes for the taxpayer, ensuring the correctness of tax software is of paramount importance. 
\emph{Metamorphic testing} has emerged as a leading solution to test and debug legal-critical tax software due to the absence of correctness requirements and trustworthy datasets.
The key idea behind metamorphic testing is to express the properties of a system in terms of the relationship between one input and its slightly metamorphosed twinned input. 
Extracting metamorphic properties from IRS tax publications is a tedious and time-consuming process. 
As a response, this paper formulates the task of generating metamorphic specifications as a \emph{translation task} between properties extracted from tax documents---expressed in natural language---to a contrastive first-order logic form. 
We perform a systematic analysis on the potential and limitations of \emph{in-context} learning with \emph{Large Language Models} (LLMs) for this task, and outline a research agenda towards automating the generation of metamorphic specifications for tax preparation software. 
\end{abstract}

\section{Introduction}

Recent surveys estimate that between 40 to 50\% of taxpayers in the United States use tax preparation software to file their taxes \cite{Farrington.2023}. Given the pervasive use of these systems and the penalties associated with filing taxes incorrectly~\cite{IRS}, making sure that tax preparation software is free of bugs is of paramount importance. 
However, there are considerable challenges that make the application of standard software testing approaches infeasible in this domain. 
Among these, the absence of correctness requirements and the lack of publicly available benchmarks (due to obvious privacy concerns) 
are primary obstacles to automatic testing and debugging of such systems.

To address these challenges, \citet{DBLP:conf/icse/TizpazNiariMWDRT23}  introduced a testing framework for U.S. tax preparation software guided by \textit{metamorphic relations}. The authors define metamorphic relations as relations between two similar individuals who differ in key characteristics that put them in different tax income buckets. This way, they are able to evaluate the outcome of an individual taxpayer in comparison with individuals who are deemed similar to them. With the help of tax preparation experts, they were able to manually derive a comprehensive set of critical correctness properties of tax preparation outcomes expressed in first-order logic, and develop a search strategy to automatically generate test cases for an open-source tax preparation software. While their approach was successful, deriving formal metamorphic specifications is a time-consuming process that requires significant domain expertise. Moreover, any changes to the tax code would require continuous efforts to make sure that specifications remain up to date. 

In this paper, we set out to explore whether recent advances in Large Language Models (LLMs) \cite{NEURIPS2020_1457c0d6, touvron2023llama, alpaca, wang-etal-2023-self-instruct} can help reduce the manual effort required to derive metamorphic specifications for this domain. 
While supervised learning without sizeable training datasets is infeasible, recent findings show that LLMs can perform surprisingly well in few-shot scenarios for a wide variety of language tasks, including traditional NLP tasks like translation and question-answering \cite{NEURIPS2020_1457c0d6}, vision-language tasks \cite{monajatipoor-etal-2023-metavl}, and socio-linguistic tasks like morality framing \cite{roy-etal-2022-towards}. 

To perform our study, we first define the extraction of metamorphic specifications for tax preparation software as a \emph{language to first-order logic translation} task. To do this, we curate a dataset of 33 high-quality natural language properties derived from tax documents\footnote{These documents include Form 1040 (U.S. Individual Income Tax Return), Publication 596 (Earned Income Tax Credit), Schedule 8812 (Qualifying Children and Other Dependents), and Form 8863 (Education Credits).} and their corresponding formal logic representation \citet{DBLP:conf/icse/TizpazNiariMWDRT23}. As curating these high-quality examples is expensive, we formulate the problem as a few-shot learning task and experiment with various \emph{in-context learning} strategies to obtain a formal representation for each property. In-context learning is a task-adaptation technique that does not update the parameters of the LLM, but rather primes the model response by providing a sequence of demonstrations \cite{NEURIPS2020_1457c0d6}. Recent work has shown that in-context learning can lead to better out-of-domain performance than few-shot fine-tuning \cite{awadalla-etal-2022-exploring,si2023prompting}. 

We perform a comprehensive analysis of our results and show that using few-shot in-context learning, the model makes between 1-2 mistakes per property on average when generating predicates, variables and operators. While this is encouraging for a model that has barely seen any in-domain, tasks-specific data, there is a lot of room for improvement before these specifications are usable for automated testing. This work represents a first step towards the automated generation of metamorphic specifications for tax preparation software. In Sec. \ref{sec:conclusion} we outline a research agenda for this purpose.



\section{Related Work}

Converting natural language utterances to logic forms has a long history in the Natural Language Processing community. Semantic parsing is a long-standing NLP task that looks at this problem \cite{kamath2019a}. Most of the work in this space has been either tailored to going from natural language to linguistically-motivated meaning representations~\cite{Palmer2010,banarescu-etal-2013-abstract}, or to executable programs like SQL queries~\cite{sun-etal-2018-semantic} and robotic commands~\cite{dukes-2014-semeval}. The solutions employed for these domains are not readily applicable to our case. On the one hand, datasets for linguistically motivated meaning representations usually deal with simple general utterances, and their structured forms are very distinct from the structured forms needed for automated software verification. On the other hand, executable domains like SQL have the advantage that there are large-scale, readily available repositories of in-domain data, which is not the case for low-resource, closed domains like tax preparation software. 

More recently, there has been increased interest in exploring NLP techniques to derive formal specifications from technical documentation. For example, \citet{DBLP:conf/sp/PachecoHWGN22} look at the task of extracting finite-state machines from network protocol RFCs, \citet{9116374} extract LTL
correctness specifications from prose policies for IoT apps, and \citet{236222} look at sentences in a developer guide to extend the finite-state machines for a small set of payment services. In all of these cases, authors are either able to leverage off-the-shelf parsers or employ supervised methods over a curated in-domain dataset.

\begin{table*}
    \centering
    \resizebox{\textwidth}{!}{%
    \begin{tabular}{ |p{10cm}|p{10cm}|  }
      \hline
      \thead{Input: Property} & \thead{Output: Specification} \\
      \hline
    \texttt{An individual with the married filing jointly 
($MFJ$) status with a disabled spouse must 
receive similar or higher tax benefits compared to a similar individual but without the disabled spouse.} &  $\forall {\bf x} ({\bf x}.sts = MFJ) \implies \forall {\bf y} (({\bf x} \equiv_{s\_blind} {\bf y}) \wedge 
({\bf x}.s\_blind \wedge \neg {\bf y}.s\_blind))
    \implies \mathcal{F}({\bf x}) \geq \mathcal{F}({\bf y})$ \\  
\hline 
    
    \end{tabular}}
    \caption{\footnotesize An example of input and expected output. Here, the quantification is over the domain of taxpaying individuals, the relation $\equiv_{pred}$ relates equal individuals except for predicate $pred$, and the function $\mathcal{F}$ characterizes the tax return for an individual.}
\label{tab:task}
\end{table*}

\section{Task Description}

Metamorphic testing is a well-studied strategy in the systems community \cite{chen2020metamorphic}. It was developed to circumvent the absence of oracles (i.e., black box modules
that decide whether the output of the system is correct for a
given input). The main idea behind metamorphic testing is that correctness can be established by contrasting multiple input-output behaviors \cite{7422146}. The running example is search-engine validation; while it might be impossible to know the expected items that should be returned for any search query $q$, if we have a more restrictive query $q'$, we know that the number of retrieved items for $q$ should be greater than the number of retrieved items for $q'$. This allows for developing strategies to test whether systems satisfy the metamorphic relations. 

To formulate our task, we build on the definition of metamorphic specifications for tax-preparation software proposed by \citet{DBLP:conf/icse/TizpazNiariMWDRT23}. We have a fixed set $P$ of metamorphic properties written in natural language, expressing aspects of the U.S. Individual
Income Tax Returns that relate to disability, credits, and deductions. For each property $p \in P$, we want to synthesize a specification $s$ in first-order logic that captures the metamorphic property. Fig. \ref{tab:task} shows an example of the input and expected output. In Appendix \ref{app:metamorphic-relations}, we detail the syntax and semantics of metamorphic specifications used for the tax preparation software, as well as the full list of formal specifications for metamorphic relations used for demonstration and testing.

\section{Experimental Evaluation}

In this section, we evaluate the performance of LLMs on the natural language to metamorphic specifications translations. We first outline the evaluation dimensions that we consider, then we describe our experimental setup and results. 

\subsection{Evaluation Dimensions}

In order to investigate LLMs' abilities to convert natural language properties into metamorphic specifications, we evaluate different dimensions to identify how the number of examples that the model gets to observe, the prompting strategies used, the LLM implementation used, as well as the characteristics of the task affect model performance. 

\textbf{Number of examples for few-shot learning:}
We look at the effect that the number of demonstrations has on model performance. Since we are working with an extremely reduced dataset, we limit the maximum number of examples that the model sees to two.

\textbf{In-context learning strategies:} We seek to study if decomposed prompting~\cite{khot2023decomposed} is effective in translating natural language into formal metamorphic specifications. To do this, we compare the performance between an end-to-end \textbf{(E2E)} prompting strategy and different decomposed strategies. To do this, we break down the task of generating metamorphic relations into two sub-tasks: (1) Generating the relevant logical predicates, and (2) Generating the specification in first-order logic. We experiment with two decomposed strategies; the first one \textbf{(Implicit)} allows the model to access the context used to generate the predicates by prompting each task subsequently, while the second one \textbf{(Explicit)} decouples the predicate and FOL generation entirely, allowing each module to completely focus on its sub-task. Example prompts for all strategies are provided in App. \ref{app:prompt_examples}. 

\textbf{LLM implementation:} We evaluate all three learning strategies on GPT-3.5. However, since we do not have access to the GPT-3.5 weights, we are reliant on APIs provided by OpenAI. In the interest of transparency and reproducibility, we also try out the Explicit strategy on Alpaca \cite{alpaca}, an open-source instruction-tuned model built over LLaMA 2 \cite{Touvron2023Llama2O}.

\textbf{Example difficulty:} Different properties can be captured by varying numbers of predicates. In our evaluation, we compare the model performance against the number of predicates required to capture a metamorphic property. We stipulate that the difficulty of capturing a metamorphic expression increases as the number of predicates increases.   

\textbf{Error types:} We qualitatively evaluate the different types of errors that the model makes. We qualify different errors by looking at specific cases where the model generated incorrect predicates, operators, or variables.





\subsection{Results}

\textbf{Experimental Setup:} For GPT-3.5, we prompt via the ChatCompletion API provided by OpenAI. We use the default API settings as outlined in their documentation. For Alpaca, we use the fine-tuned weights from the LoRa-adapted \cite{Hu2021LoRALA} Alpaca 7B model, and run inference to generate responses for our prompts.

\textbf{Performance Metrics and Annotation:} Standard generation metrics like BLEU, ROGUE, ChrF and BLEURT rely either on word and character matching \cite{papineni-etal-2002-bleu, lin-2004-rouge, popovic-2015-chrf}, or semantic embedding similarity \cite{sellam-etal-2020-bleurt}. Metrics that rely on lexical matching are ill-suited to evaluate the generation of first-order-logic propositions, as the operators, predicate names and variables used can vary significantly while maintaining semantic consistency (e.g., the operation $x \equiv_{S} y$ and the predicate $EqualExcept(x,y, S)$ express the same relationship). Similarly, embedding-based metrics that have been trained on large textual repositories do not account for valid substitutions in logic that lack similarity in natural language (e.g., $\mathcal{F}(x)$ and $TaxReturn(x)$ can be used to denote the same predicate). For this reason, we resort to human evaluation and define a quality score ranging from 1 to 5 based on the following rubric: 

\begin{table}[H]
\centering
\resizebox{\columnwidth}{!}{%
\begin{tabular}{@{}cl@{}}
\toprule
Rating & \multicolumn{1}{c}{Explanation} \\ \midrule
5 & The generated predicates or FOL match the ground truth. \\
4 & The generated identities have the correct semantic sense, but incorrect format. \\
3 & There is 1 mistake in the set of {predicates, variables, operators} \\
2 & There are 2 mistakes in the set of {predicates, variables, operators} \\
1 & The generated identities are completely incorrect. \\ \bottomrule
\end{tabular}%
}
\label{tab:rubric}
\caption{Rubric for Performance Quality}
\end{table}

\begin{table*}[ht]
\resizebox{\textwidth}{!}{%
\begin{tabular}{@{}ccccccccc@{}}
\toprule
\multirow{2}{*}{\textbf{n-shot}} & \multicolumn{2}{c}{\textbf{GPT E2E}} & \multicolumn{2}{c}{\textbf{GPT Implicit}} & \multicolumn{2}{c}{\textbf{GPT Explicit}} & \multicolumn{2}{c}{\textbf{Alpaca Explicit}} \\ \cline{2-9} 
                                 & \textbf{Pred.}     & \textbf{FOL}    & \textbf{Pred. }       & \textbf{FOL}      & \textbf{Pred.}        & \textbf{FOL}       & \textbf{Pred.}         & \textbf{FOL}         \\ \cline{1-1}
0                                & 1.77 (0.99)        & 1.19 (0.49)     & 1.15 (0.46)           & 1.03 (0.19)       & 1.00 (0.00)          & 1.03 (0.19)        & 1.00 (0.00)           & 1.00 (0.00)          \\
1                                & 3.29 (1.38)        & 2.25 (1.48)     & 3.46 (1.39)           & 1.27 (0.60)       & 4.03 (1.02)          & 2.77 (1.28)        & 1.61 (1.09)           & 1.74 (2.02)          \\
2                                & 3.96 (1.15)        & 2.85 (1.70)     & 3.65 (1.09)           & 1.42 (0.95)       & 4.27 (0.87)          & 3.16 (1.23)        & 1.63 (1.08)           & 1.71 (1.16)          \\ \hline
\end{tabular}%
}
\caption{Average (S.D.) quality scores for Predicate and FOL generation by in-context learning strategy.}
\label{tab:nshot_vs_prompt}
\end{table*}

\begin{table*}[]
\centering
\begin{tabular}{@{}ccccccc@{}}
\toprule
\multicolumn{1}{l}{\multirow{2}{*}{\textbf{Strategy}}} & \multicolumn{2}{c}{\textbf{\# Predicates $<$ 4}} & \multicolumn{2}{c}{\textbf{4 $\le$ \# Predicates $<$ 6}} & \multicolumn{2}{c}{\textbf{\# Predicates $\ge$ 6}} \\ \cmidrule(l){2-7} 
\multicolumn{1}{l}{}                                   & \textbf{Pred.}    & \textbf{FOL}    & \textbf{Pred.}     & \textbf{FOL}     & \textbf{Pred.}     & \textbf{FOL}    \\ \midrule
GPT E2E                                                    & 3.83 (1.60)                           & 3.16 (2.04)                     & 4.07 (0.95)                             & 3.31 (1.54)                       & 3.88 (1.25)                             & 1.88 (1.46)                      \\
GPT Implicit                                               & 3.50 (1.22)                           & 1.83 (1.60)                      & 3.66 (0.88)                             & 1.50 (0.79)                       & 3.75 (1.39)                            & 1.00 (0.00)                     \\
GPT Explicit                                               & 4.20 (1.03)                           & 3.50 (0.97)                     & 4.58 (0.67)                            & 3.58 (1.08)                       & 3.88 (0.83)                             & 2.13 (1.25)                      \\ \bottomrule
\end{tabular}

\caption{Average (S.D.) quality scores for Predicate and FOL generation for each in-context learning strategy, grouped by difficulty level in terms of the number of ground truth predicates in the statement.}\label{tab:num_preds}
\end{table*}


Using this rubric, the first two authors of this paper annotated the generated texts with their qualitative judgments. A total of 33 examples were evaluated (see Appendix \ref{app:metamorphic-relations}), and average results are reported in Tables \ref{tab:nshot_vs_prompt} and \ref{tab:num_preds}.


\textbf{Predicate and FOL quality based on prompting style and number of examples:} In Table \ref{tab:nshot_vs_prompt}, we look at the effect that examples have on different styles of prompting. We found that on average, \textbf{GPT Explicit} decomposed prompting performed best among all strategies when demonstrations were provided. Additionally, we find higher variation in results for \textbf{GPT E2E}, suggesting that decomposed prompting might lead to slightly more consistent results. As expected, demonstrating with a higher number of examples helps the model generate better translations. Nevertheless, the average quality for all strategies sits solidly in the middle of the rubric (about 1-2 errors per example), shedding light on the limitations of few-shot in-context learning as-is for this task. \textbf{Alpaca Explicit} produces nearly identical results to \textbf{GPT Explicit} in the zero-shot scenario. However, performance does not improve when Alpaca is provided with two examples instead of just one, unlike \textbf{GPT Explicit}. The Alpaca implementation tested is significantly smaller (7B vs. 175B for GPT 3.5). We hypothesize that Alpaca 7B likely needs access to more context to be able to improve its performance for this task. We leave an in-depth exploration of the capabilities of smaller LLMs for future work.



\textbf{Effect of predicate complexity on translation:} In this evaluation, we fix the number of examples the model sees to two. We then evaluate how different numbers of predicates affect the quality of translations. Results can be seen in Tab. \ref{tab:num_preds}. For all strategies, there is a clear drop in performance for harder examples. This is consistent with our hypothesis that more complex propositions are more difficult to translate. Similarly to the previous analysis, we find that performance on decomposed prompting performs better than E2E prompting. Given the overhead of manual evaluation, we limited this test to GPT-3.5. 


In Appendix \ref{app:example_rubric}, we provide generation examples for each score in the annotation rubric. 


\subsection{Qualitative Analysis of Errors}

In this section, we qualitatively evaluate different errors after qualifying them on the basis of correctness with regard to 1) Predicates; 2) Variables; and 3) Operators. We observe that while the model is somewhat robust in translating natural language into logic, it encounters problems when trying to express identities in metamorphic forms. 

\textbf{Errors observed in predicates:} The most common error that we observed in the translations was due to a lack of consideration for predicates that are expected in metamorphic templates; such as $EqualExcept(x,y, S)$. where $x$ and $y$ are entities that are similar except through the set of predicates defined in $S$.

\textbf{Errors observed in variables:} We see that models often declare variables or omit variables without much structure. Due to the stochastic nature of LLM predictions, it often reuses  (or is primed to reuse) the same variables it has declared for different predicates; automatically invalidating the metamorphic translation. 

\textbf{Errors observed in operators:} Models often imitate the same operators that were provided in the prompts. For instance, when we provided a prompt that used $\geq$; the prompt generated the prediction using the same operator to describe the opposite kind of relationship. Similarly, the model frequently uses commonly occurring pairs of logic notation like ( $\forall $ and $\exists$).

\textbf{Summarizing qualitative analysis:} Our qualitative study revealed that while LLMs hold the potential to perform well on metamorphic translation tasks, we also observe simple errors caused due to stochastic dependencies. More training data or post-hoc filtering may benefit this approach from generating erroneous responses.

\section{Findings, Limitations and Future Work}\label{sec:conclusion}

Our experiments show that while LLMs are able to make some headway in the task of metamorphic translation, more work needs to be done to improve performance in order to build safeguards for tax preparation software. Below, we list three key directions for future work. 

\textbf{Task-oriented prompt learning:} It is clear that few-shot in-context learning is not enough to fully automate the generation of metamorphic specifications for tax preparation software. Recent work has shown that learning prompts that are tailored specifically to the task at hand can help improve in-context learning performance \cite{chung2022scaling, han-etal-2022-meet, zhang-etal-2023-assisting}. Going forward, we would like to explore strategies to dynamically adapt prompting strategies to maximize performance and minimize common mistakes. 

\textbf{Leveraging additional data sources:} While we do not have access to a comprehensive dataset containing metamorphic specifications, there are several external resources that we could indirectly exploit to design weakly-supervised methods and representation learning strategies for the task at hand. For example, we could exploit unlabeled tax-centric documents, out-of-domain language-to-logic datasets, and LLMs trained over code and other structured representations. There is a body of work on task-adaptive pre-training~\cite{gururangan-etal-2020-dont}, domain adaptation~\cite{daume-iii-2007-frustratingly}, and code-based LLMs for semantic parsing~\cite{shin-van-durme-2022-shot} that we could exploit to either complement or replace vanilla in-context learning. 

\textbf{Closing the loop:} While we explored the task of generating metamorphic specifications by translating from properties in natural language to contrastive first-order logic forms, this task alone is not sufficient to fully automate the verification of tax preparation software. To close the loop, two tasks remain: (1) Automatically extracting properties from free-form tax documents, and (2) Taking potentially noisy specifications and generating executable tests. While closing the loop is an ambitious task, we hypothesize that these additional steps could help inform the translation process, potentially improving performance. For example, if we could automatically extract more properties -even if noisy- from raw documents, we could easily expand the training examples without additional manual cost. Moreover, connecting the extracted specifications with symbolic, executable modules, could serve as a form of indirect feedback to inform the translation module.
\section{Acknowledgements}

Tizpaz-Niari and Trivedi were partially supported by the NSF under grants
CCF-2317206 and CCF-2317207. This work utilized the Alpine high-performance computing resource at the University of Colorado Boulder. Alpine is jointly funded by the University of Colorado Boulder, the University of Colorado Anschutz, Colorado State University, and the NSF (award 2201538).

\bibliography{emnlp2023-latex/anthology, emnlp2023-latex/custom}
\bibliographystyle{emnlp2023-latex/acl_natbib}

\appendix
\section{Metamorphic Relations.}
\label{app:metamorphic-relations}

\noindent\textbf{Syntax and Semantics.} We review the syntax and semantics of first-order logic for metamorphic specification in the context of tax preparation software. 
Let $X = \set{X_1, X_2, \ldots, X_n}$ is the set of variables corresponding to various fields
about an individual in the tax return form and $\mathcal{F}: \mathcal{D}_1 {\times} \mathcal{D}_2 {\times} \cdots {\times} \mathcal{D}_n {\to} \Real_{\geq 0}$ is the \emph{federal tax return} computed by the software,
where $\mathcal{D}_i$ is the domain of variable $X_i$.
We write $\mathcal{D}$ for $\mathcal{D}_1 \times \mathcal{D}_2 \times \cdots \times \mathcal{D}_n$.
These variables correspond to intuitive labels such as ${\tt age}$ (numerical variable),
${\tt blind}$ (Boolean variable), and ${\tt sts}$ (filing status with values such as MFJ, married filing jointly, and MFS, married filing separately). 
For an individual ${\bf x} \in \Dd$, we write ${\bf x}(i)$ for the value of $i$-th variable,
or ${\bf x}.{\tt lab}$ for the value of variable ${\tt lab}$.
Let $\mathcal{L}$ be the set of all labels.

For labels $L \subseteq \mathcal{L}$ and inputs ${\bf x} \in \mathcal{D}$ and ${\bf y} \in \mathcal{D}$, we say that ${\bf y}$ is a metamorphose of ${\bf x}$ with the exceptions of labels $L$, and we write ${\bf x} \equiv_{L} {\bf y}$ if 
$\forall \ell \not \in L \text{ we have that } {\bf x}.\ell = {\bf y}.\ell$.
A metamorphic relation is a first-order logic formula with variables in $X$, constants from domains in $\mathcal{D}$, relation $\equiv_L$, comparisons $\set{<, \leq, = , \geq, >}$  over numeric variables, predicate $\neg$ (negation) for Boolean valued labels, real-valued function for federal tax return $\mathcal{F}: \Dd \to \Real$, Boolean connectives $\wedge$, $\lor$, $\neg$, $\implies$, $\iff$, and quantifiers $\exists x. \phi(x)$ and $\forall x. \phi(x)$ with natural interpretations. 
We assume that the formulas are given in the prenex normal form, i.e. a block of quantifiers followed by a quantifier free formula.

\vspace{0.5em}
\noindent\textbf{Selected Specifications.} Following \cite{DBLP:conf/icse/TizpazNiariMWDRT23}, we consider aspects of Individual Income Tax Return that relate to disability, credits, and deductions. Specifically, we focus on on fields related to the standard deductions for senior and disable individuals; the Earned Income
Tax Credit (EITC), a refundable tax credits for lower income households; the Child Tax Credit (CTC), a non-refundable credit to reduce the taxes owed based on the number of qualifying children under the age of 17; the Educational Tax Credit (ETC) that helps students with the cost of higher education by lowering their owed taxes or increasing their refund; and the Itemized Deduction (ID), an option for taxpayers with significant tax deductible expenses.

We use scenarios and examples described in the policies above to synthesize metamorphic relations. Table~\ref{tab:metamorphic-relations} shows $33$ metamorphic relations in $6$ domains for the tax year 2020. For properties
\#9 to \#12, we assume $MAGI$ (modified adjusted gross income) is equivalent to $AGI$.
Next, we provide a brief explanation of some of these properties. 

\begin{enumerate}
    \item A senior (over age of 65) must receive similar or better tax benefits when compared to a person without the
    seniority who is similar in every other aspect (due to higher standard deductions for seniors).

    \item A blind individual must receive similar or better tax benefits when compared to a person without the
    disability who is similar in every other aspect (due to higher standard deductions for blind individuals).

    \item An individual with the married filing jointly ($MFJ$) status
    with a senior spouse must receive similar or higher tax benefits compared to a similar individual but without
    the senior spouse.

    \item An individual with the married filing jointly ($MFJ$) status
    with a disabled spouse must receive similar or higher tax benefits compared to a similar individual but without
    the disabled spouse.

    \item An individual who files with the head of household ($HoH$) status
    should receive similar or higher tax return benefits compared to a similar individual who
    files with the single status. 

    \item An individual who files the tax with the qualified widow ($QW$) status
    should receive similar or higher tax return benefits compared to a similar individual who
    files the tax with single status. 

    \item An individual who files the tax with the qualified widow ($QW$) status
    should receive similar or higher tax return benefits compared to a similar individual who
    files with the head of household ($HoH$) status. 
        
    \item An individual with the married filing separately ($MFS$) status
    who claims EITC credits should receive the same tax return compared
    to a similar individual (with the same status) who does not claim EITC credits. 

    \item An individual with the married filing jointly ($MFJ$) status with 
    $AGI$ over $56,844$ who claims EITC credits should receive the same tax return compared
    to a similar individual (with the same status) who does not claim EITC credits. 
    
    \item An individual with the married filing jointly ($MFJ$) status with 
    $AGI$ less than or equal $56,844$ who claims EITC credits should receive a higher tax returns compared
    to a similar individual who has $AGI$ greater than $56,844$.
    
    \item Among two qualified
    individuals with EITC, one with higher EITC claims receives higher or equal tax return benefits.
    
    \item An individual who has the investment income less than
    \$3,650, computed as the sum of lines 2a, 2b, 3a, and 7 in Form 1040, should receive 
    a higher tax returns compared to a similar individual with the investment income greater than
    \$3,650.
    
    \item An individual with single status with $AGI$ less than or equal to $15,820$ who is $25$ years old or older
    should receive a higher tax returns compared to a similar individual who is younger than $25$ years old.
    
    \item An individual with head of household status with $AGI$ less than or equal to $15,820$ who is $25$ years old or older should receive a higher tax returns compared to a similar individual who is younger than $25$ years old.
    
    \item An individual with head of household status with $AGI$ less than or equal to $41,757$ and with
    one qualified children who is younger than $25$ years old should receive a similar return compared to
    a similar individual who is $25$ years or older (the age test is not relevant when having at least
    one qualified children).
    
    \item An individual with qualifying widow status with $AGI$ less than or equal to $15,820$ who is $25$ years old or older without any qualified children should receive a higher tax returns compared to a similar individual who is younger than $25$ years old.
    
    \item An individual with Qualifying Widow status with $AGI$ less than or equal to $41,757$ and with
    one qualified children who is younger than $25$ years old should receive a similar return compared to
    a similar individual who is $25$ years or older.
    
    \item  A married filing jointly status with $AGI$ less than or equal to $21,710$ who
    is $25$ years old or older without any qualified children should receive a higher tax returns compared to a similar individual who is younger than $25$ years old.
    
    \item An individual with Qualifying Widow status with $AGI$ less than or equal to $47,646$ and with
    one qualified children who is younger than $25$ years old should receive a similar return compared to
    a similar individual who is $25$ years or older.

    \item An individual who is qualified for EITC credit with no qualified children should receive
    a similar or lower tax returns compared to a similar individual with one qualified children.

    \item An individual who is qualified for EITC credit with one or less qualified children should receive
    a similar or lower tax returns compared to a similar individual with two qualified children.

    \item An individual who is qualified for EITC credit with two or less qualified children should receive
    a similar or lower tax returns compared to a similar individual with three qualified children.

    \item An individual who is qualified for EITC credit with three or less qualified children should receive
    a similar tax returns compared to a similar individual with more than three qualified children.
    
    \item Among two qualified
    married filing jointly ($MFJ$) individuals, one with higher child tax credits receives higher or equal  tax return benefits.
    
    \item This 4-property requires a comparison between
    four ``similar'' individuals since there is a relation between two variables of interests:
    $AGI$ and the number of qualified children/others to claim a CTC.
    An individual with more qualified dependents
    must receive higher or similar tax return benefits than an individual with fewer dependents after adjusting for the effects of income levels on the calculations of both the final return and the amounts of CTC claims. Expressing this property requires holding the income of two individuals the same per each qualified number of children/others.
    
    \item An individual with the married filing separately ($MFS$) status who claims
    CTC credits should receive similar tax benefits compared to a similar individual
    with $MFS$ status who does not claim CTC. 

    \item  An individual with the married filing jointly ($MFJ$) status
    with $AGI$ over $180k$ who claims ETC should receive similar tax benefits compared to a similar individual
    who does not claim ETC. 
    
    \item An individual with the married filing jointly ($MFJ$) status with $AGI$ below $160k$ who claims ETC received higher or similar tax return benefits compared to a similar individual who does not claim
    ETC or claims a lower amount of ETC credit.

    \item
    This 4-property requires a comparison between four
    ``similar'' individuals as the rule changes for individuals with $AGI$ below $160$k and between $160$k and $180$k.
    By holding $AGI$ constant between two individuals with $AGI$ below $160$k (varying the $ETC$ claims) and 
    two individuals with $AGI$ between $160$k and $180$k (varying
    the $ETC$ claims with the same rate), the property requires that individuals with lower income (below $160k$) receive higher or similar
    tax returns.

    \item An individual who files
    with medical/dental expenses ($MDE$) below 7.5\% of their $AGI$ and itemizes their
    deductions receives the same return as a similar individual with no $MDE$ claims.
    
    \item An individual who files with a standard deduction who claims 
    itemized deduction (Line $12$) should receive similar tax returns compared
    to a similar individual who does not claim itemized deduction.

    \item An individual who files 
    with itemized deductions below the standard deductions receive a lower or similar
    tax return benefits compared to a similar individual who files with the standard deductions.

    \item An individual who files 
    with itemized deductions above the standard deductions receive a higher or similar
    tax return benefits compared to a similar individual who claims standard deductions.
\end{enumerate}

\begin{table*}[]
    \caption{Metamorphic properties for five domains in the US tax (2020) policies.
    $\mathcal{F}$ is federal tax return where negative
    values mean the individual owns payment to the IRS, $sts$ is filing status, $s\_lab$ is spouse's field $lab$,
    $MFJ$: married filing jointly, $MFS$ is married filing separately, $S$ is single filing, $HoH$ is head of household filing, $QW$ is qualifying widow filing, $AGI$ is adjusted gross income, $L27$ is line $27$ of IRS $1040$ for Earned Income Tax Credit (EITC), $QC$ is the number of qualified children, $OD$ is the number of other dependents, $CTC$ is child tax credits,
    $L19$ is line $19$ of IRS $1040$ for Child Tax Credit ($CTC$),
    $L29$ is line $29$ of IRS $1040$ for Education Tax Credit ($ETC$), $MDE$ is medical/dental expenses reported in line $1$ of schedule $A$, $iz$ is to use itemized deductions ($ID$) vs. standard deductions, and $L12$ is total itemized deductions ($ID$) from schedule $A$.}
    \label{tab:metamorphic-relations}
    \centering
    \resizebox{\textwidth}{!}{
    \begin{tabularx}{\textwidth}{@{} l l L @{}}
     \hline
     Id & Domain & Metamorphic Property \\
     \hline
    1  & Disability & $\forall {\bf x}, {\bf y} (({\bf x} {\equiv_{age}} {\bf y}) \wedge ({\bf x}.age {\geq} 65) \wedge ({\bf y}.age {<} 65 ))
    \implies \mathcal{F}({\bf x}) \geq \mathcal{F}({\bf y})$ \\
     \hline
     2  & Disability & $\forall {\bf x}, {\bf y} (({\bf x} {\equiv_{blind}} {\bf y}) \wedge ({\bf x}.blind \wedge \neg {\bf y}.blind))
    \implies \mathcal{F}({\bf x}) \geq \mathcal{F}({\bf y})$ \\
     \hline    
     3  & Disability &  $\forall {\bf x} ({\bf x}.sts{=}MFJ) {\implies} \forall {\bf y} (({\bf x} \equiv_{s\_age} {\bf y}) \wedge ({\bf x}.s\_age{\geq}65) \wedge ({\bf y}.s\_age{<}65))
    {\implies} \mathcal{F}({\bf x}) \geq \mathcal{F}({\bf y})$ \\
     \hline    
     4  & Disability &  $\forall {\bf x} ({\bf x}.sts = MFJ) \implies \forall {\bf y} (({\bf x} \equiv_{s\_blind} {\bf y}) \wedge ({\bf x}.s\_blind \wedge \neg {\bf y}.s\_blind))
    \implies \mathcal{F}({\bf x}) \geq \mathcal{F}({\bf y})$ \\    
    \hline
    5  & Status & $\forall {\bf x}, {\bf y} (({\bf x} {\equiv_{sts}} {\bf y}) \wedge ({\bf x}.sts {=} HoH) \wedge ({\bf y}.sts {=} S))
    \implies \mathcal{F}({\bf x}) \geq \mathcal{F}({\bf y})$ \\
     \hline
    6  & Status & $\forall {\bf x}, {\bf y} (({\bf x} {\equiv_{sts}} {\bf y}) \wedge ({\bf x}.sts {=} QW) \wedge ({\bf y}.sts {=} S))
    \implies \mathcal{F}({\bf x}) \geq \mathcal{F}({\bf y})$ \\
     \hline
    7  & Status & $\forall {\bf x}, {\bf y} (({\bf x} {\equiv_{sts}} {\bf y}) \wedge ({\bf x}.sts {=} QW) \wedge ({\bf y}.sts {=} HoH))
    \implies \mathcal{F}({\bf x}) \geq \mathcal{F}({\bf y})$ \\
     \hline     
     8  & EITC  & $\forall {\bf x} ({\bf x}.sts = MFS) \implies \forall {\bf y} ({\bf x} {\equiv_{L27}} {\bf y} \wedge {\bf x}.L27 >0.0 \wedge {\bf y}.L27 = 0.0) \implies \mathcal{F}({\bf x}){=}\mathcal{F}({\bf y})$ \\
     \hline
     9  & EITC  & $\forall {\bf x} ({\bf x}.sts{=}MFJ) \wedge ({\bf x}.AGI{>}56,844)\implies \forall {\bf y} ({\bf x} {\equiv_{L27}} {\bf y} \wedge {\bf x}.L27{>}0.0 \wedge {\bf y}.L27{=}0.0) \implies \mathcal{F}({\bf x}){=} \mathcal{F}({\bf y})$ \\
     \hline
     10  & EITC & $\forall {\bf x} (({\bf x}.sts{=}MFJ)\wedge({\bf x}.L27{>}0.0)){\implies}\forall {\bf y} ({\bf x} {\equiv_{AGI}} {\bf y} \wedge {\bf x}.AGI{\leq}56,844 \wedge {\bf y}.AGI{>}56,844) {\implies}\mathcal{F}({\bf x}){\geq} \mathcal{F}({\bf y})$ \\
     \hline
     11  & EITC & $\forall {\bf x} ({\bf x}.sts{=}MFJ){\wedge}({\bf x}.AGI{\leq}56,844)
     {\implies}$ 
     $\forall {\bf y} (({\bf x} {\equiv_{L27}} {\bf y}){\wedge}{\bf x}.L27{\geq}{\bf y}.L27){\implies}\mathcal{F}({\bf x}){\geq}\mathcal{F}({\bf y})$ \\ 
     \hline
     12  & EITC  & $\forall {\bf x} ({\bf x}.L27{>} 0.0) \forall {\bf y} ({\bf x} {\equiv_{\{L2a,L2b,3b,7\}}} {\bf y} \wedge {\bf x}.L2a + {\bf x}.L2b + {\bf x}.L3b + {\bf x}.L7 > 3,650) \wedge {\bf y}.L2a + {\bf y}.L2b + {\bf y}.L3b + {\bf y}.L7 {\leq} 3,650) \implies \mathcal{F}({\bf x}){\leq}\mathcal{F}({\bf y})$ \\
     \hline     
     13  & EITC & $\forall {\bf x} ({\bf x}.sts{=}S){\wedge}({\bf x}.AGI{\leq}15,820){\wedge}({\bf x}.L27{>} 0.0)
     {\implies} \forall {\bf y} (({\bf x} {\equiv_{Age}} {\bf y}){\wedge}({\bf x}.Age {<} {25}) \wedge ({\bf y}.Age \geq {25})) \implies \mathcal{F}({\bf x}){<}\mathcal{F}({\bf y})$ \\ 
     \hline
     14  & EITC & $\forall {\bf x} ({\bf x}.sts{=}HoH){\wedge}({\bf x}.AGI{\leq}15,820){\wedge}({\bf x}.L27{>} 0.0){\wedge}({\bf x}.QC{=} 0)$
     ${\implies} \forall {\bf y} (({\bf x} {\equiv_{Age}} {\bf y}){\wedge}({\bf x}.Age {<} {25}) \wedge ({\bf y}.Age \geq {25})) \implies \mathcal{F}({\bf x}){<}\mathcal{F}({\bf y})$\\ 
     \hline
     15  & EITC & $\forall {\bf x} ({\bf x}.sts{=}HoH){\wedge}({\bf x}.AGI{\leq}41,756){\wedge}({\bf x}.L27{>} 0.0){\wedge}({\bf x}.QC{=} 1) {\implies}$
     $\forall {\bf y} (({\bf x} {\equiv_{Age}} {\bf y}){\wedge}({\bf x}.Age{<}{25}) \wedge ({\bf y}.Age \geq {25})) \implies \mathcal{F}({\bf x}){=}\mathcal{F}({\bf y})$ \\ 
     \hline
     16  & EITC & $\forall {\bf x} ({\bf x}.sts{=}QW){\wedge}({\bf x}.AGI{\leq}15,820){\wedge}({\bf x}.L27{>} 0.0){\wedge}({\bf x}.QC{=} 0){\implies}$
     $\forall {\bf y} (({\bf x} {\equiv_{Age}} {\bf y}){\wedge}({\bf x}.Age{<}{25}) \wedge ({\bf y}.Age \geq {25})) \implies \mathcal{F}({\bf x}){<}\mathcal{F}({\bf y})$ \\ 
     \hline
     17  & EITC & $\forall {\bf x} ({\bf x}.sts{=}QW){\wedge}({\bf x}.AGI{\leq}41,756){\wedge}({\bf x}.L27{>} 0.0){\wedge}({\bf x}.QC{=} 1) {\implies}$
     $\forall {\bf y} (({\bf x} {\equiv_{Age}} {\bf y}){\wedge}({\bf x}.Age{<}{25}) \wedge ({\bf y}.Age \geq {25})) \implies  \mathcal{F}({\bf x}){=}\mathcal{F}({\bf y})$ \\ 
     \hline     
     18  & EITC & $\forall {\bf x} ({\bf x}.sts{=}MFJ){\wedge}({\bf x}.AGI{\leq}21,710){\wedge}({\bf x}.L27{>} 0.0){\wedge}({\bf x}.QC{=} 0) {\implies}$
     $\forall {\bf y} (({\bf x} {\equiv_{Age}} {\bf y}){\wedge}({\bf x}.Age{<}{25}) \wedge ({\bf y}.Age \geq {25})) 
     \implies \mathcal{F}({\bf x}){<}\mathcal{F}({\bf y})$ \\ 
     \hline          
     19  & EITC & $\forall {\bf x} ({\bf x}.sts{=}MFJ){\wedge}({\bf x}.AGI{\leq}47,646){\wedge}({\bf x}.L27{>} 0.0){\wedge}({\bf x}.QC{=} 1) {\implies}$
     $\forall {\bf y} (({\bf x} {\equiv_{Age}} {\bf y}){\wedge}({\bf x}.Age{<}{25}) \wedge ({\bf y}.Age \geq {25})) \implies \mathcal{F}({\bf x}){=}\mathcal{F}({\bf y})$\\      
     \hline
     20  & EITC & $\forall {\bf x} ({\bf x}.L27{>} 0.0){\wedge}({\bf x}.QC{=} 0) {\implies}\forall {\bf y} (({\bf x} {\equiv_{QC}} {\bf y}){\wedge}({\bf y}.QC{=}{1}) \implies \mathcal{F}({\bf x}){\leq}\mathcal{F}({\bf y})$ \\      
     \hline          
     \end{tabularx}
     }
\end{table*}

\begin{table*}[!h]
\caption*{Continuation of Previous Table.}
\centering
    \resizebox{\textwidth}{!}{
    \begin{tabularx}{\textwidth}{@{} l l L @{}}
     \hline
     Id & Domain & Metamorphic Property \\
     \hline
     21  & EITC & $\forall {\bf x} ({\bf x}.L27{>} 0.0){\wedge}({\bf x}.QC{\leq} 1) {\implies}\forall {\bf y} (({\bf x} {\equiv_{QC}} {\bf y}){\wedge}({\bf y}.QC{=}{2}) \implies \mathcal{F}({\bf x}){\leq}\mathcal{F}({\bf y})$ \\      
     \hline          
     22  & EITC & $\forall {\bf x} ({\bf x}.L27{>} 0.0){\wedge}({\bf x}.QC{\leq} 2) {\implies}\forall {\bf y} (({\bf x} {\equiv_{QC}} {\bf y}){\wedge}({\bf y}.QC{=}{3}) \implies \mathcal{F}({\bf x}){\leq}\mathcal{F}({\bf y})$ \\      
     \hline          
     23  & EITC & $\forall {\bf x} ({\bf x}.L27{>} 0.0){\wedge}({\bf x}.QC{=} 3) {\implies}\forall {\bf y} (({\bf x} {\equiv_{QC}} {\bf y}){\wedge}({\bf y}.QC{>}{3}) \implies \mathcal{F}({\bf x}){=}\mathcal{F}({\bf y})$ \\      
     \hline          
     24  & CTC & $\forall {\bf x} ({\bf x}.sts{=}MFS){\wedge}({\bf x}.AGI{\leq}200k){\implies}$
     $\forall{\bf y} (({\bf x} {\equiv_{L19}} {\bf y}){\wedge}({\bf x}.L19{\geq}{\bf y}.L19)){\implies}\mathcal{F}({\bf x}){\geq}\mathcal{F}({\bf y})$)\\   
     \hline 
     25  & CTC & $\forall {\bf x}, {\bf x'} ({\bf x}.sts {=} {\bf x'}.sts {=} MFJ) 
     {\wedge}({\bf x}.AGI {<} 400k){\wedge}({\bf x'}.AGI {\geq} 400k) 
     {\wedge}{\lceil}{\bf x'}.AGI{-}400k{\rceil}_{1k}*0.05{<} {\bf x'}.QC*2k+{\bf x}.OD*0.5k 
     {\implies} 
     \forall {\bf y}, {\bf y'} ({\bf x} {\equiv_{\{QC,OD\}}} {\bf y}){\wedge}({\bf x'} {\equiv_{\{QC,OD\}}} {\bf y'})  
     \wedge  (0 {\leq} {\bf y}.QC {=} {\bf y'}.QC {\leq} {\bf x}.QC {=} {\bf x'}.QC \leq 10)
     \wedge  (0 {\leq} {\bf y}.OD {=} {\bf y'}.OD \leq {\bf x}.OD {=} {\bf x'}.OD \leq 10)
     \implies (\mathcal{F}({\bf x}) - \mathcal{F}({y})) \geq (\mathcal{F}(x') - \mathcal{F}(y'))$\\       
     \hline
     26  & ETC  & $\forall {\bf x} ({\bf x}.sts = MFS) \implies \forall {\bf y} ({\bf x} {\equiv_{L29}} {\bf y} \wedge {\bf x}.L29 >0.0 \wedge {\bf y}.L29 = 0.0) \implies \mathcal{F}({\bf x}){=}\mathcal{F}({\bf y})$ \\
     \hline
     27 & ETC  & $\forall {\bf x} ({\bf x}.sts{=}MFJ) \wedge ({\bf x}.AGI{\geq}180k) \implies \forall {\bf y} ({\bf x} {\equiv_{L29}} {\bf y} \wedge {\bf x}.L29{>}0.0 \wedge {\bf y}.L29{=}0.0){\implies}\mathcal{F}({\bf x}){=} \mathcal{F}({\bf y})$ \\
     \hline
     28 & ETC  & $\forall {\bf x} ({\bf x}.sts{=}MFJ) \wedge ({\bf x}.AGI{\leq}160k) \implies \forall {\bf y} ({\bf x} {\equiv_{L29}} {\bf y} \wedge {\bf x}.L29{\geq}{\bf y}.L29){\implies}\mathcal{F}({\bf x}){\geq} \mathcal{F}({\bf y})$ \\
     \hline
     29  & ETC & $\forall {\bf x}, {\bf x'} ({\bf x}.sts {=} {\bf x'}.sts {=} MFJ) 
     \wedge ({\bf x}.AGI {\leq} 160k) \wedge (160k{<}{\bf x'}.AGI {<} 180k) 
     \implies
     \forall {\bf y}, {\bf y'} (({\bf x} \equiv_{L29} {\bf y}) \wedge ({\bf x'} \equiv_{L29} {\bf y'}) \wedge ({\bf x}.L29={\bf x'}.L29 \geq {\bf y}.L29={\bf y'}.L29)) 
     {\implies}(\mathcal{F}({\bf x}) - \mathcal{F}({y})) \geq (\mathcal{F}(x') - \mathcal{F}(y'))$\\       
     \hline
     30  & ID & $\forall {\bf x},{\bf y} ({\bf x} {\equiv_{MDE}} {\bf y}) \wedge ({\bf x}.MDE{\leq}{\bf x}.AGI*7.5\%)  \wedge ({\bf y}.MDE{=}0.0) \implies\mathcal{F}({\bf x}){=}\mathcal{F}({\bf y})$ \\
     \hline 
     31  & ID & $\forall {\bf x} (\neg {\bf x}.iz) \implies \forall {\bf y} ({\bf x} {\equiv_{MDE}} {\bf y} \wedge {\bf x}.MDE{>}0.0 \wedge{\bf y}.MDE{=}0.0) \implies\mathcal{F}({\bf x}){=}\mathcal{F}({\bf y})$ \\
     \hline
     32  & ID & $\forall {\bf x} ({\bf x}.sts{=}MFJ){\implies}$
     
     $\forall {\bf y} (({\bf x} {\equiv_{iz,L12}} {\bf y}){\wedge}({\bf x}.iz{\wedge}{\neg}{\bf y}.iz){\wedge}({\bf x}.L12{\leq}24.8k{\wedge}{\bf y}.L12{=}0.0)){\implies} \mathcal{F}({\bf x}){\leq}\mathcal{F}({\bf y})$ \\
     \hline    
     33  & ID & $\forall {\bf x} ({\bf x}.sts{=}MFJ){\implies}$
     
     $\forall {\bf y} (({\bf x} {\equiv_{iz,L12}} {\bf y}){\wedge}({\bf x}.iz{\wedge}{\neg}{\bf y}.iz){\wedge}({\bf x}.L12{>}24.8k{\wedge}{\bf y}.L12{=}0.0)){\implies} \mathcal{F}({\bf x}){\geq}\mathcal{F}({\bf y})$ \\
     \hline     
     \end{tabularx}%
     }
\end{table*}

\onecolumn
\section{Prompt Examples}\label{app:prompt_examples}

\begin{figure}[H]

    \centering
    \caption{Example of an End2End prompt, that looks at generating predicates and FOL in the same prompt. This is an example for nshot = 1; the example was omitted and another added for 0-shot and 2-shot respectively. }
    \includegraphics[width=0.6\columnwidth]{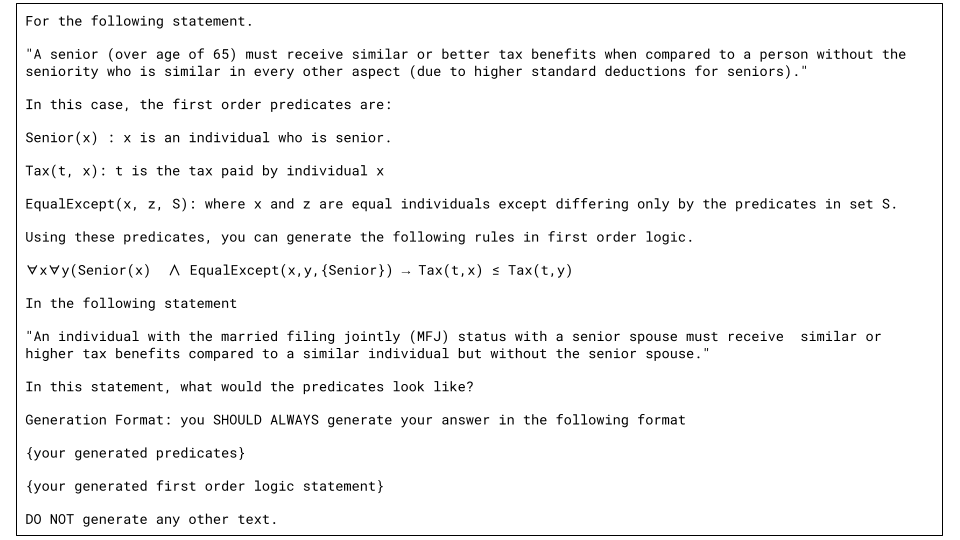}

    \vspace{0.5cm}

    \caption{Example of a prompt to generate predicates. The prompt is the same for explicit and implicit setups. This is an example for nshot = 1; the example was omitted and another added for 0-shot and 2-shot respectively. }
    \includegraphics[width=0.6\columnwidth]{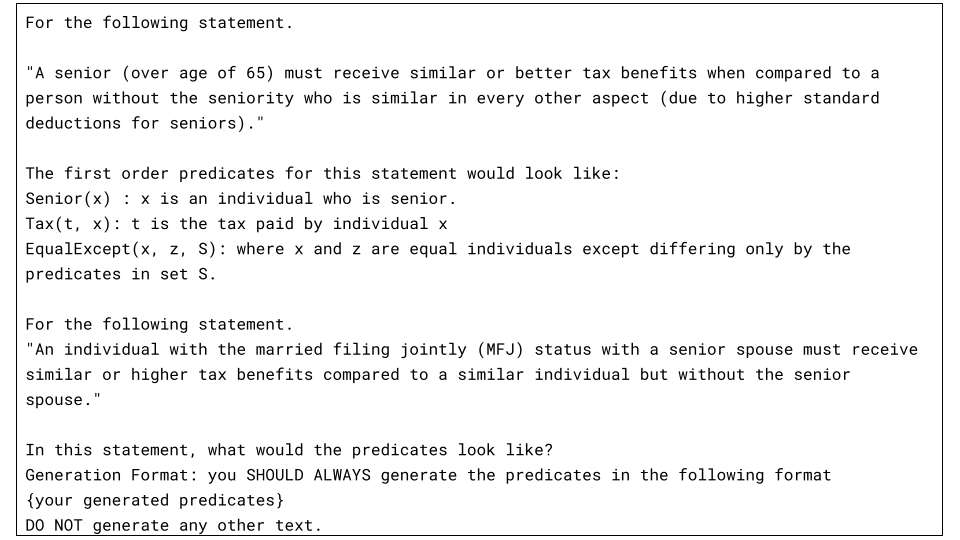}

    \vspace{0.5cm}

    \caption{Example of a prompt to generate FOL, given that predicates were already generated. This setup is only used in explicit cases (as predicates are provided through context in implicit cases). This is an example for nshot = 1; the example was omitted and another added for 0-shot and 2-shot respectively. }
    \includegraphics[width=0.7\columnwidth]{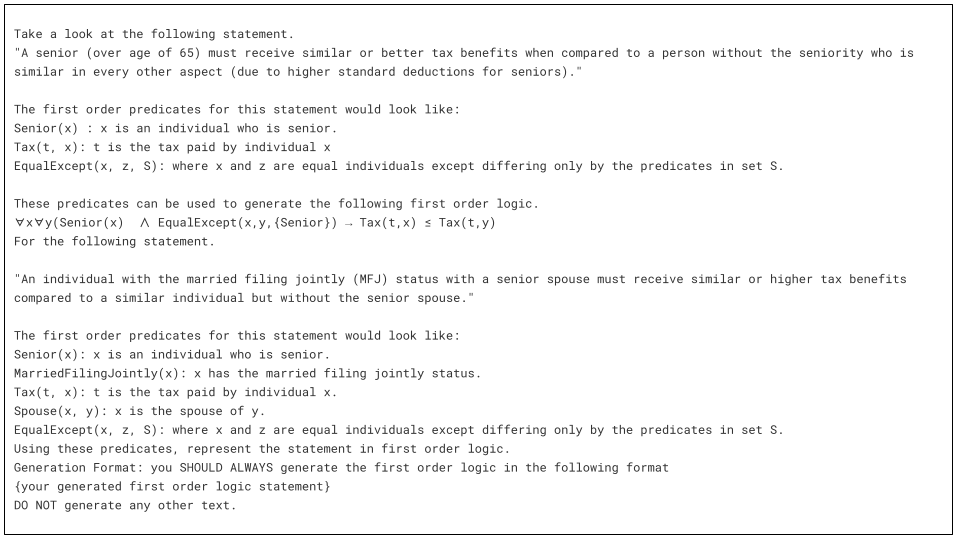}
    
\end{figure}

\section{Example Generations for Annotation Rubric}\label{app:example_rubric}

\begin{table*}
\centering
\resizebox{\textwidth}{0.90\textheight}{%
    \Rotatebox{90}{%
        \begin{tabular}{|l|l|l|l|l|}
            \hline
            Metamorphic Statement & Generated Predicates & Generated First Order Logic & Explanation & Score \\ \hline
            \begin{tabular}[c]{@{}l@{}}An individual with the married filing jointly (MFJ) status \\ with AGI over \$56,844 who claims EITC credits should \\ receive the same tax return compared to a similar \\ individual (with the same status) who does not claim \\ EITC credits.
            \end{tabular} 
            & \begin{tabular}[c]{@{}l@{}}MFJ(x): x is an individual with the \\ married filing jointly status.\\ AGI(y, x): y is the adjusted gross \\ income of individual x.\\ ClaimEITC(x): x claims EITC credits.\\ TaxReturn(t, x): t is the tax return of \\ individual x.\\ EqualExcept(x, z, S): where x and z \\ are equal individuals except\\ differing only by the \\ predicates in set S.
            \end{tabular} 
            & \begin{tabular}[c]{@{}l@{}}$\forall$x$\forall$y(MFJ(x) $\land$ AGI(y,x) $>$ 56844 $\land$ \\ ClaimEITC(x) $\land$ \\ EqualExcept(x,y,{ClaimEITC}) \\ $\implies$ TaxReturn(t,x) $=$ TaxReturn(t,y)
            \end{tabular} & 
            \begin{tabular}[c]{@{}l@{}}The model correctly predicts that\\  AGI criteria is applicable to both \\ individuals in question, and only \\ differs on EITC status.
            \end{tabular}
            & 5 \\ \hline
            \begin{tabular}[c]{@{}l@{}}An individual with single status with AGI less than or equal to \\ \$15,820 who is 25 years old or older should receive a higher \\ tax returns compared to a similar individual who is younger \\ than 25 years old.
            \end{tabular} 
            & \begin{tabular}[c]{@{}l@{}}Single(x): x is an individual with single status.\\ AGI(x, a): x has adjusted gross income of a.\\ Younger(x, y): x is younger than y years old.\\ TaxReturns(r, x): x receives r in tax returns.\end{tabular} 
            & \begin{tabular}[c]{@{}l@{}}$\forall$x$\forall$y(Single(x) $\land$ AGI(x, a) $\leq$ 15820 $\land$ Older(x, 25) \\ $\land$ Younger(y, 25) $\land$ EqualExcept(x,y,{AGI}) $\implies$ \\ TaxReturns(r, x) $>$ TaxReturns(r, y)
            \end{tabular} 
            & \begin{tabular}[c]{@{}l@{}}All the predicates have been \\ generated correctly. However, \\ the predicate "EqualExcept" \\ should only apply to age of \\ the individual and not the AGI. \\ This makes some semantic \\ sense, but is incorrect (not \\ usable).
            \end{tabular} 
            & 4 \\ \hline
            \begin{tabular}[c]{@{}l@{}}An individual who files the tax with the qualified widow (QW) \\ status should receive similar or higher tax return benefits \\ compared to a similar individual who files the tax with single \\ status.
            \end{tabular} 
            & \begin{tabular}[c]{@{}l@{}}QW(x): x is an individual who files the tax with \\ the qualified widow status.\\ Single(x): x is an individual who files the tax \\ with single status.\\ TaxReturn(t, x): t is the tax return received by \\ individual x.\\ SimilarOrHigher(x, y): individual x receives \\ similar or higher tax return benefits \\ compared to individual y.
            \end{tabular} 
            & \begin{tabular}[c]{@{}l@{}}$\forall$x$\forall$y(QW(x) $\land$ EqualExcept(x,y,{QW}) $\implies$ \\ SimilarOrHigher(x,y)
            \end{tabular} 
            & \begin{tabular}[c]{@{}l@{}}The system makes a mistake in \\ predicates - it does not use the \\ "Single" status to make a \\ judgement on tax return status.
            \end{tabular} 
            & 3 \\ \hline
            \begin{tabular}[c]{@{}l@{}}An individual with the married filing jointly (MFJ) status with a \\ disabled spouse must receive similar or higher tax benefits \\ compared to a similar individual but without the disabled spouse.\end{tabular} 
            & \begin{tabular}[c]{@{}l@{}}MFJ(x): x is an individual with the married filing \\ jointly status.\\ Disabled(y): y is an individual who is disabled.\\ Tax(t, x): t is the tax paid by individual x.\\ SimilarOrHigher(x, z, S): where x and z are similar \\ individuals except for the predicates in set S.
            \end{tabular} 
            & \begin{tabular}[c]{@{}l@{}}$\forall$x$\forall$y(MFJ(x) $\land$ Disabled(y) $\land$ \\ SimilarOrHigher(x, y, {Disabled}) $\implies$ \\ Tax(t, x) $\leq$ Tax(t, y)
            \end{tabular} 
            & \begin{tabular}[c]{@{}l@{}}Here, the system misses the \\ predicate "Spouse(x)". In \\ addition to this, the greater than \\ operator is flipped (or there is a \\ mistake in the "Tax(t, x)" predicate \\ as well).
            \end{tabular} 
            & 2 \\ \hline
            \begin{tabular}[c]{@{}l@{}}An individual with the married filing jointly (MFJ) status with a \\ senior spouse must receive similar or higher tax benefits \\ compared to a similar individual but without the senior spouse.
            \end{tabular} 
            & \begin{tabular}[c]{@{}l@{}}MFJ(x): x is an individual with the married filing \\ jointly status.\\ SeniorSpouse(y): y is a senior spouse of an \\ individual.\\ Tax(t, x): t is the tax paid by individual x.\\ EqualExcept(x, z, S): where x and z are equal \\ individuals except differing only \\ by the predicates in set S.
            \end{tabular} 
            & \begin{tabular}[c]{@{}l@{}}$\forall$x$\forall$y(MFJ(x) $\land$ SeniorSpouse(y) $\land$ \\ EqualExcept(x, y, {MFJ, SeniorSpouse}) $\implies$ \\ Tax(t, x) $\leq$ Tax(t, y)
            \end{tabular} 
            & \begin{tabular}[c]{@{}l@{}}Here, the SeniorSpouse is incorrectly \\ applied to individual y. Additionally, \\ the predicate EqualExcept takes more \\ than one predicate (and is redundant). \\ Lastly, the $\leq$ is flipped (or the \\ "Tax" predicate is incorrect).
            \end{tabular} 
            & 1 \\ \hline
        \end{tabular}
    }
}
\caption{In this table, we show examples of our qualitative score rubric. A rating of $5$ usually means that the generated logic can be used for metamorphic testing. All these examples were sourced from 2-shot examples in the explicit setup.}
\end{table*}

\end{document}